\documentclass[twocolumn,prb,amssymb,showpacs,superscriptaddress]{revtex4}

\usepackage{dcolumn}
\usepackage{bm}

\usepackage{graphicx}
\usepackage{color}
\newcommand{\field}[1]{\mathbb{#1}}

\newcommand{\bk}{{\bf k}}
\newcommand{\etilde}{\tilde{\epsilon}}

\begin{document}
\title{Tunneling-driven breakdown of the 331 state and the emergent Pfaffian and composite Fermi liquid phases}
\author{Z. Papi\'{c}}
\affiliation{Scientific Computing Laboratory, Institute of Physics, University of Belgrade, P.~O.~Box 68, 11 000 Belgrade, Serbia}
\affiliation{Laboratoire de Physique des Solides, Univ.~Paris-Sud, CNRS, UMR 8502, F-91405 Orsay Cedex, France}
\affiliation{Laboratoire Pierre Aigrain, Ecole Normale Sup\'erieure, CNRS, 24 rue Lhomond, F-75005 Paris, France}
\author{M. O. Goerbig}
\affiliation{Laboratoire de Physique des Solides, Univ.~Paris-Sud, CNRS, UMR 8502, F-91405 Orsay Cedex, France}
\author{N. Regnault}
\affiliation{Laboratoire Pierre Aigrain, Ecole Normale Sup\'erieure, CNRS, 24 rue Lhomond, F-75005 Paris, France}
\author{M. V. Milovanovi\'c}
\affiliation{Scientific Computing Laboratory, Institute of Physics, University of Belgrade, P.~O.~Box 68, 11 000 Belgrade, Serbia}

\begin{abstract} 

We examine the possibility of creating the Moore-Read Pfaffian in the lowest Landau level when the multicomponent Halperin 331 state (believed to describe quantum Hall bilayers and wide quantum wells at the filling factor $\nu=1/2$) is destroyed by the increase of tunneling. Using exact diagonalization of the bilayer Hamiltonian with short-range and long-range (Coulomb) interactions
in spherical and periodic rectangular geometries, we establish that tunneling is a perturbation that drives the 331 state into a compressible composite Fermi liquid, with the possibility for an intermediate critical state that is reminiscent
of the Moore-Read Pfaffian. These results are interpreted in the two-component BCS model for Cauchy pairing with a tunneling constraint. We comment on the conditions to be imposed on a system with fluctuating density in order to achieve the stable Pfaffian phase. 
\end{abstract}


\maketitle \vskip2pc

\section{Introduction} \label{sec_introduction}

When electrons are confined to a two-dimensional (2D) plane and subject to a strong perpendicular magnetic field, they
organize themselves into fascinating strongly-correlated quantum phases. The most prominent examples are the 
Laughlin states,\cite{laughlin} $|\Psi_{\mathrm L} \rangle = \Phi_{2k+1}(\{ z \})$, which may be written in terms of the 
Laughlin-Jastrow factor $\Phi_{m}(\{ z \})=\prod_{i<j}(z_i-z_j)^m,$ where $z_j=x_j+iy_j$ denotes the complex coordinate of the
$j$-th electron, $m,k$ are integers and we have neglected the Gaussian factor ubiquitous in the lowest Landau level (LLL). 
Laughlin states describe the fractional quantum Hall effect (FQHE) that occurs when the filling factor $\nu=N/N_{\phi}$, which is defined as the ratio between the number of electrons $N$ and the number of flux quanta $N_{\phi}$ threading the 2D system, is a simple fraction with an odd denominator, $\nu=1/(2k+1)$.\cite{TSG} Laughlin's construction has been generalized, within the framework of the composite-fermion (CF) theory,\cite{jain} to explain the rich phenomenology of a whole sequence of observed odd denominator fractions. According to the CF theory,
one understands the FQHE as an integer quantum Hall effect in an effective magnetic field that vanishes at $\nu=1/2$. As a consequence, CFs then form a compressible CF Fermi liquid (CFL), \cite{rr, hlr} $|\Psi_{\mathrm{FL}}\rangle = \mathcal{P}_{LLL} \det{\left[ \phi_i (z_j) \right]} \Phi_{2}(\{ z \}),$ as seen in the absence of the Hall plateau in single thin layers at $\nu=1/2$ in the LLL.\cite{jiang} Here, as before, the characteristic quantum Hall correlations are captured in the Jastrow factor $\Phi_{2}(\{ z \})$ which we refer to as the \emph{charge} part of the wave function (as it carries the flux through the system) and $\phi_i(z_j)$ are the single particle states (an overall projection  $\mathcal{P}_{LLL}$ to the LLL may be needed to yield a proper trial wave function that is analytic in $z_j$.). 

However, some quantized \emph{even} denominator states exist\cite{willet} and are usually associated with the first excited 
Landau level, where the nature of the effective interaction is believed to facilitate the pairing between CFs.\cite{rh} 
The paradigm of such paired states is the so-called ``Pfaffian'' state introduced by Moore and Read,\cite{mr} 
$|\Psi_{\mathrm{Pf}}\rangle = \mathrm{Pf}\left[1/(z_i-z_j)\right] \Phi_{2}(\{ z \})$, which explains the FQHE 
observed at $\nu=5/2$.\cite{willet} In addition to the charge part $ \Phi_{2}(\{ z \})$, which fixes the filling factor, we have also a pairing in the \emph{neutral} sector described by the object $\mathrm{Pf}.$\cite{mr} In contrast to the Laughlin and Jain states with anyonic excitations satisfying Abelian statistics, 
the Moore-Read state represents the simplest paired state of spin-polarized electrons which supports excitations with 
non-Abelian statistics\cite{readrezayi} of interest in topologically protected quantum computation.\cite{tqc}

If the spin of the electrons is not necessarily frozen out by the magnetic field, the electrons could find it more favorable 
to reorganize themselves into one of the competing Abelian phases called multicomponent Halperin states.\cite{halperin} 
In these states the Hall quantization is a result of internal degrees of freedom of the electrons (provided by the spin or layer index). At half filling, a two-component candidate is the 331 state, $|\Psi_{331} \rangle = \Phi_{3}^{\mathrm{intra}}( \{ z_\uparrow \}) 
\Phi_{3}^{\mathrm{intra}}(\{ z_\downarrow \}) \Phi_{1}^{\mathrm{inter}}(\{ z_\uparrow,z_\downarrow \}),$ 
written as a straightforward generalization of the Laughlin state to two species of electrons $\uparrow$ and 
$\downarrow$. 
In order to satisfy the constraint of the fixed filling factor $\nu=1/2$, apart from the usual Laughlin-Jastrow factors between the 
electrons of the same species ($\Phi^{\mathrm{intra}}$), one also has to account for the inter-species correlations through 
the factor  $\Phi_{k}^{\mathrm{inter}}(\{ z , w \})=\prod_{i,j}(z_i-w_j)^k$. Alternatively, using the Cauchy identity\cite{hr_spinsinglet} 
we can cast the 331 state into the form 
$|\Psi_{331}\rangle = \det \left[ 1/( z_{i,\uparrow}-z_{j,\downarrow}) \right] \Phi_{2}(\{ Z \})$ 
which extracts the charge part $\Phi_2$ ($Z$'s denote all particles regardless of their spin index) from the neutral 
part where the pairing is described in terms of a Cauchy determinant between $\uparrow$ and $\downarrow$ particles. 
Numerical calculations\cite{he, ymg, y2, papic_onefourth, dassarma_bilayer} indicate that Halperin's 331 wave function is likely to be at the origin of the $\nu=1/2$ FQHE in bilayer quantum Hall systems\cite{eisen1/2, suen_bilayer} as well as in wide quantum wells.\cite{suen}

In this paper we investigate whether it is possible to create the Moore-Read Pfaffian in the LLL by converting 
the two-component 331 state into a single-component state. Mathematically, this is easily achieved by antisymmetrizing the 
neutral (Cauchy determinant) part of the $|\Psi_{331}\rangle$ between $\uparrow$ and $\downarrow$.\cite{Cappelli01,RGJ} 
However, such a procedure is a very complex mathematical entity because it creates a state with different 
physical properties (non-Abelian statistics out of the Abelian), while we are interested in a physical mechanism that 
mimics the antisymmetrization in an experimental situation. We restrict the discussion to the Coulomb bilayer system, 
which is a generic two-component system where the ``spin'' $\sigma = \uparrow,\downarrow$ denotes the two layers in which 
the electrons are localized. In such a system, it is commonly speculated that the antisymmetrization mechanism
is provided by the tunneling term $\sim -\Delta_{SAS} S_x,$ which combines the single-particle wave functions into 
symmetric (even) and antisymmetric (odd) superpositions, $\uparrow \pm \downarrow$. The tunneling term favors the even superposition (channel) where one expects to find a weakly paired (Moore-Read) phase. We establish 
that this route towards the Moore-Read state is complicated by the presence of the compressible CFL, which is the resulting 
phase for large tunneling. Along the way, one may arrive at a critical state that shares some properties with the Moore-Read 
Pfaffian, but we do not find evidence that this state represents a stable phase. 
Recent experiments\cite{shabani, luhman} found even denominator fractions in wide quantum wells in the LLL. 
The results obtained within the present bilayer model may be relevant also in the study of wide quantum wells insofar
as the latter can, with moderate approximations, effectively be described by a bilayer Hamiltonian, where the tunneling term mimics 
the effective confinement gap between the lowest and the first excited electronic sublevel.\cite{papic_onefourth}

The remainder of this paper is organized as follows. In Sec. \ref{sec_tunneling} we introduce the BCS model for spinful fermions with tunneling, first proposed by Read and Green.\cite{read_green}
This model decouples into an even and odd channel, and the tunneling term increases the population of the even channel, 
where one expects to find the weak pairing (Pfaffian) phase. However, the increase of tunneling also leads to the effective 
weakening of the coupling, which eventually drives the system into a CFL phase when the total number of particles is held 
fixed. This is the situation we analyze in detail in our exact diagonalization calculations in spherical and toroidal geometries 
(Sec. \ref{sec_numerics}). These results are discussed in the context of the phase diagram of Ref. \onlinecite{read_green}
(see Sec. \ref{sec_discussion}). We furthermore introduce a generalized tunneling constraint which, in the BCS description, 
leads to a stable weak pairing (Pfaffian) phase in the even channel when the density of the of the system is not fixed 
(Sec. \ref{sec_generalizedtunneling}). We present our conclusions in Sec.\ref{sec_conclusions}.

\section{BCS model with tunneling}\label{sec_tunneling}

At $\nu=1/2$, the CFs experience a zero net magnetic field,\cite{jiang} and if we limit ourselves to the neutral part of $|\Psi_{\mathrm{Pf}}\rangle$, they may be described within the framework of the effective BCS model introduced in Ref.
\onlinecite{read_green}. We consider the system to be at zero temperature and neglect fluctuations in the Chern-Simons 
gauge field that are related to the charge part of $|\Psi_{\mathrm{Pf}}\rangle$.\cite{bonesteel} The Hamiltonian which
describes the Cauchy pairing between $\uparrow$ and $\downarrow$ particles with tunneling $\Delta_{SAS}$ reads
\begin{eqnarray} \label{ham}
\nonumber H &=& \sum_{\bk} \left[\etilde_{\bk} (c_{\bk \uparrow}^{\dagger}c_{\bk \uparrow} + c_{\bk
\downarrow}^{\dagger}c_{\bk \downarrow}) +  \left(\Delta_{\bk} c_{\bk
\uparrow}^{\dagger}c_{-\bk \downarrow}^{\dagger}+\, {\rm H.c.}\right)
\right.\\
 &&\left. - \frac{\Delta_{SAS}}{2} (c_{\bk \uparrow}^{\dagger}c_{\bk
\downarrow}+c_{\bk \downarrow}^{\dagger}c_{\bk \uparrow})\right], 
\end{eqnarray}
where $\etilde_{\bk} = \epsilon_{\bk} - \mu$, in terms of the putative CF dispersion relation $\epsilon_{\bk}$ and the 
chemical potential $ \mu$, which is assumed positive $\mu>0$. Notice that because of the vanishing net magnetic field, the
2D wave vector $\bk=(k_x,k_y)$ is again a good quantum number. The order parameter $\Delta_{\bk}=\Delta_0(k_x - ik_y)$
is chosen to describe $p$-wave pairing, and we assume that $\Delta_{\bk}$ and $\mu$ are not renormalized by the tunneling. 

With the help of the even, $c_{\bk,e}=(c_{\bk,\uparrow} +c_{\bk,\downarrow})/\sqrt{2}$, and odd spin combinations 
$c_{\bk,o}=(c_{\bk,\uparrow} - c_{\bk,\downarrow})/\sqrt{2}$, the Hamiltonian (\ref{ham}) decouples into an even and odd channel,\cite{read_green} $H=H^e + H^o$, where (the index $\tau$ denotes the even and odd channel, $\tau=e,o$)
\begin{equation}\label{hamEO}
H^{\tau} = \sum_{\bk}\left[(\epsilon_{\bk} - \mu^{\tau})c_{\bk,\tau}^{\dagger}c_{\bk,\tau} + \left(\Delta_{\bk}^{\tau}c_{\bk,\tau}^{\dagger}c_{-\bk,\tau}^{\dagger} +\, {\rm H.c.}\right)
\right],
\end{equation}
in terms of the chemical potentials $\mu^e=\mu +\Delta_{SAS}/2$ and $\mu^o=\mu - \Delta_{SAS}/2$ for the even and odd channels,
respectively. Furthermore, the even/odd $p$-wave order parameters read $\Delta_{\bk}^e = \Delta_{\bk}/2= (\Delta_0/2)(k_x - ik_y)$ 
and $\Delta_{\bk}^o = \Delta_{-\bk}/2= -(\Delta_0/2)(k_x - ik_y)$.

For moderate tunnelings, the effective chemical potential $\mu^{\rm eff}$ of the whole system may be viewed
as the weighted sum of the two channels, $\mu^{\rm eff} = P \mu^e + (1-P)\mu^o,$ where $P$ measures the population of 
the even channel ($1/2 \leq P \leq 1$) and may have a complicated dependence on $\Delta_{SAS}$. 
In particular, for some values of $\Delta_{SAS}$ we may be below the critical line $\mu^{\rm eff}=P \Delta_{SAS}$ and 
inside the non-Abelian (Pfaffian) phase. However, in the limit of large tunneling, the system is dominated by the
even channel and the chemical potential of the whole system is $\mu^{\rm eff} = \mu^e$ because $P=1$.  
Remember that the associated BCS wave function in the even channel reads
\begin{equation}\label{pairingFon}
|\psi_{BCS}\rangle = \prod_{\bk}\left( 1 + g_{\bk}c_{\bk,e}^{\dagger}c_{-\bk,e}^{\dagger}\right)|{\rm vacuum}\rangle,
\end{equation}
in terms of the pairing function $g_{\bk}\sim v_{\bk}/u_{\bk} \sim \mu^e/ \Delta_0$. One notices then that
an increase of the chemical potential $\mu^e$ controlled by the large value of the tunneling parameter 
$\Delta_{SAS}$ is equivalent to a reduction of the order parameter $\Delta_{0}$. Therefore the BCS system will eventually be transformed into the one of the Fermi liquid. We can see this more explicitly by examining the relevant excitations of the even channel,\cite{read_green}
\begin{equation}
E = \sqrt{(\epsilon_{\bk} - \mu^e)^{2} + \Delta_{0}^{2} k^{2}},
\end{equation}
in the limit of large $\mu^e$ around $k=|\bk| = 0$. They become unstable and $\bk = 0$ becomes a point of local maximum. The minimum
is expected to move to $ |\bk| = k_{F}$, the Fermi momentum.\cite{read_green} Therefore if $\Delta_{0}$ does not ``renormalize"
with increasing $\Delta_{SAS}$, the net effect of the strong tunneling ($\mu^{e} \gg \mu$) on the Cauchy pairing is to drive the
system into a Fermi liquid. This is not unexpected because one retrieves an effective one-component system in this limit, where all
particles are ``polarized`` in the even channel. The Pfaffian physics may however play a role in the intermediate state before 
complete polarization. We revisit the BCS approach in Sec. \ref{sec_generalizedtunneling}, with a slightly different perspective 
in which the antisymmetrization is imposed,  in a functional formalism, with the help of a Lagrangian multiplier which plays a 
similar role as the present tunneling term $\Delta_{SAS}$.

As we pointed out earlier, the population of the even channel $P$ may be a complicated function of tunneling. In the following 
section we use exact diagonalization of small finite systems in order to get a hint of the form of this dependence 
$P=P(\Delta_{SAS})$ and determine the nature of possible phases as $P$ increases from $1/2$ to $1$.

\section{Exact diagonalization}\label{sec_numerics}

Here we study the full interacting quantum Hall bilayer Hamiltonian for small finite systems in the presence of tunneling,\cite{y2, dassarma_bilayer}

\begin{eqnarray}\label{main_hamiltonian}
\nonumber H &=& - \Delta_{SAS} S_x + \sum_{i<j, \sigma \in \uparrow, \downarrow} V^{\mathrm{intra}}(\mathbf{r}_{i\sigma}-\mathbf{r}_{j\sigma}) \\
&& + \sum_{i,j} V^{\mathrm{inter}}(\mathbf{r}_{i\uparrow} - \mathbf{r}_{j\downarrow}),
\end{eqnarray}
where in coordinate representation we have 
$2S_x=\int d\mathbf{r} \left[ \Psi_\uparrow^\dagger (\mathbf{r}) \Psi_\downarrow(\mathbf{r}) + {\rm H.c.}\right]$,  
$\Psi_{\sigma}^\dagger(\mathbf{r})$ creates a particle at the position $\mathbf{r}$ in the layer $\sigma$.  
We have decomposed the interaction into terms between electrons belonging to the same layer ($V^{\mathrm{intra}}$) 
and those residing in opposite layers ($V^{\mathrm{inter}}$). We consider a short-range interaction, defined as 
\begin{equation}\label{shortrange}
V_{331}^{\mathrm{intra}}(r)=V_1\nabla^2\delta(r), \qquad V_{331}^{\mathrm{inter}}(r)=V_0\delta(r),
\end{equation}
which produces the $331$ state as the densest and unique zero energy state when $V_0,V_1$ are chosen positive.\cite{hierarchy, readrezayi} We also consider long-range Coulomb interaction, 
\begin{equation}\label{longrange}
V_{\rm Coul}^{\mathrm{intra}}(r)=e^2/\epsilon r,\qquad V_{\rm Coul}^{\mathrm{inter}}(r)=e^2/\epsilon \sqrt{r^2+d^2},
\end{equation}
where $d$ is the distance between layers. We fix the total number of particles in our calculations to be an even integer and, unless stated otherwise, take $d=l_B$ ($l_B$ is the magnetic length), which merely sets the range for the distance between the layers where the Coulomb ground state is supposed to be fairly well described by the 331 wave function. Confining the electrons to a compact surface such as the sphere\cite{hierarchy} or torus,\cite{haldane_torus} the Hilbert space becomes finite and one may exactly diagonalize the interacting Hamiltonian (\ref{main_hamiltonian}).\cite{haldane_rezayi_ed} The ground state obtained in this way can be numerically compared with the trial wave functions $|\Psi_{331}\rangle$ and $|\Psi_{\mathrm{Pf}}\rangle$ as a simple scalar product between vectors in the Hilbert space. In these calculations $|\Psi_{\mathrm{Pf}}\rangle$ is defined in the even basis i.e. single-particle states are understood to be even combinations of the original $\uparrow, \downarrow$ states. 

\subsection{Sphere}\label{sec_sphere}

If we wrap the electron sheet into a sphere and place a magnetic monopole in the center which generates $N_\phi$ magnetic flux quanta perpendicular to the surface, we are left with a finite basis of single particle states indexed by $0,...,N_\phi$. Translational symmetry in the plane is replaced by rotational symmetry, which leads to a classification of the many-body states by the eigenvalues of angular momentum $L$ and its $z$-component $L_z$.\cite{hierarchy} The two-body interaction such as (\ref{shortrange}) and (\ref{longrange}) is parametrized by Haldane pseudopotentials $V_L^{\sigma\sigma'}$ which represent the energy of a pair of particles located in layers $\sigma,\sigma'$ with relative angular momentum $L$.\cite{hierarchy} Incompressible quantum Hall states are invariably obtained in the $L=0$ sector of the Hilbert space. They are further characterized by a topological number called the \emph{shift}\cite{hierarchy} $\delta$, defined through the relation $N_\phi = \frac{1}{\nu}N - \delta$. To specify uniquely the Hilbert space corresponding to the given trial state, one needs to know $\delta$ to get the correct value for the pair of $(N,N_\phi)$. Sometimes it is possible to find different values of $\nu, \delta$ that yield the same $(N, N_\phi)$ -- this is called the aliasing problem because two different trial states get realized in the same Hilbert space. We disregard such cases in our calculations.

\begin{figure}[htb]
\centering
\includegraphics[width=\textwidth, angle=270, scale=0.3]{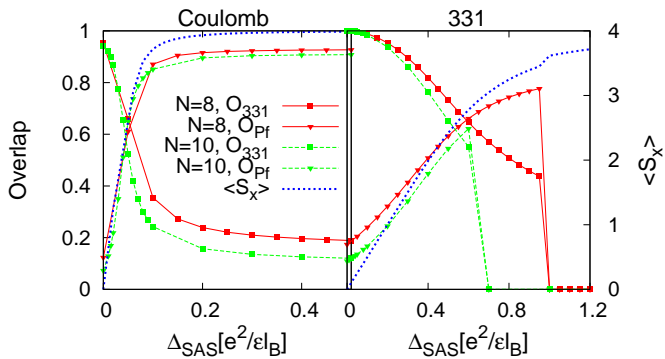}
\caption{(Color online) Overlaps between the exact ground state of the Coulomb bilayer (left panel) and short range 331 Hamiltonian (right panel) with the 331 state ($O_{331}$) and the Pfaffian ($O_{\mathrm{Pf}}$), as a function of tunneling $\Delta_{SAS}$. Also shown on the right axis is the expectation value of the $S_x$ component of pseudospin (for $N=8$ system) which characterizes the two-component to one-component transition.}
\label{fig_sphere}
\end{figure}

In Fig.\ref{fig_sphere} we present results of exact diagonalization on the sphere for the short-range (331) Hamiltonian (\ref{shortrange}) and long-range (Coulomb) Hamiltonian (\ref{longrange}). 331 and Pfaffian trial states occur at the same value of the shift, thus we are able to track their evolution as a function of tunneling simultaneously. We also use $\langle S_x \rangle$, the expectation value of the $S_x$ operator in the ground state, to monitor the two-component to one-component transition, whereas $\langle S_z \rangle$ remains zero throughout, which is due to the weaker interlayer as compared to the intralayer interaction. Starting from the long-range Hamiltonian (Fig. \ref{fig_sphere}, left panel), we see that the 331 state gives way to a Pfaffian-like ground state, with the overlap quickly saturating to a value of around $0.92$.\cite{y2, dassarma_bilayer} The transition occurs for $\Delta_{SAS}\simeq 0.1e^2/\epsilon l_B$ which agrees well with the typical experimental value and shows little size dependence when the largest accessible system $N=10$ is considered (note that the subsequent $N=12$ system suffers from the aliasing problem). On the other hand, notice that for the short-range Hamiltonian (Fig. \ref{fig_sphere}, right panel), the 331 state
is much more robust to perturbation by $\Delta_{SAS}$: before it reaches full polarization in the $x$-direction (maximum $\langle S_x \rangle$), the overlap with both incompressible states drops precipitously beyond some critical $\Delta_{SAS}$ which is rather size-sensitive (it also depends on the values of the parameters one chooses in Eq. (\ref{shortrange}), but the qualitative features of the transition are reproduced for many different choices of $V_0,V_1$).

We see nonetheless that the breakdown of a two-component phase yields a one-component state manifested by $\langle S_x \rangle \rightarrow N/2$ (this is the limit $P \rightarrow 1$ from Sec.\ref{sec_tunneling}).  At the transition, $\langle S_x \rangle$ develops a small but visible kink. Focusing on the large tunneling limit, we find that the nature of the ground state is effectively that of the single layer (polarized) ground state for the symmetric interaction $V^+(r)=\left[V^{\mathrm{intra}}(r) + V^{\mathrm{inter}}(r) \right]/2$. This intuitive result was directly verified for all the available system sizes, including very large $N=10$ system on both sphere and torus (see Sec. \ref{sec_torus}). In view of this, it is not surprising that the large-tunneling limit of the short-range Hamiltonian is the compressible CFL: $V^+$ in this case reduces to the repulsive hard core $V_1$ pseudopotential which has a tendency to produce the Jain CF state. This is also apparent in the fact that the ground states for large tunneling are obtained in the angular momentum sectors that agree with the predictions for the excitations of the CFL yielding overlap of $0.99$ with the excited CF sea ground state.

Therefore, the results for the short-range Hamiltonian are suggestive that we may have a direct 331-CFL transition in the thermodynamic limit because the transition point seems to be shifting towards smaller tunnelings as we increase $N$. 
Notice, however, that in contrast to the incompressible 331 and Pfaffian states, which occur at a shift $\delta=3$, the CFL has
a shift $\delta=4$. On the sphere, the two incompressible states can thus not be directly compared to the CFL, and the evidence
for the 331-CFL transition is therefore indirect. This problem is circumvented in ED on the torus presented hereafter in Sec. \ref{sec_torus}. In the Coulomb case, on the other hand, we observe a curious saturation of the ground state overlap with the Pfaffian. We attribute this feature to the effect of the long range Coulomb potential on a finite system. One notices that by adding an asymptotic tail to the ``intra" component of the short-range pseudopotentials $V_L^{\mathrm{intra}}=V_{L, 331}^{\mathrm{intra}}+\alpha/2\sqrt{L}$ ($\alpha$ nonzero for $L\geq 3$) , one progressively increases the critical value of $\Delta_{SAS}$ for the abrupt drop of the overlaps as $\alpha \rightarrow 1$ (pure Coulomb). In fact, for $N=8,10$ it is sufficient to consider only $V_3^{\mathrm{intra}}$ to achieve the saturation and push the critical value of $\Delta_{SAS}$ to infinity. 

Results for the Coulomb interaction in the large $\Delta_{SAS}$ limit (Fig. \ref{fig_sphere}) are similar to those obtained in Ref.\onlinecite{papic_zds} where single-layer Zhang-Das Sarma interaction was used. As long as we are in the large $\Delta_{SAS}$ limit, $V^+(r)$ interaction produces numerically the same effect as the Zhang-Das Sarma interaction. In particular, transition to a Moore-Read Pfaffian will be induced if the layer separation $d$ is sufficiently large.\cite{papic_zds} Of course, these two interactions are different from each other and the fact that they yield the same phenomenology (phase transitions as $d$ is varied) only means we are probing a critical state where even the slightest perturbation away from pure Coulomb interaction (coupled with the bias of the shift) is sufficient to generate incompressibility. However, despite large overlap, the gap remains very small after the transition. The difference between the two interactions is obvious in the torus geometry (Sec.\ref{sec_torus}). The new result of the present paper is that we find the Pfaffian signature even in the region without full polarization ($P \lesssim 1$), as we elaborate in Sec.\ref{sec_discussion}.

\subsection{Torus}\label{sec_torus}

Another way to compactify an infinite plane is to impose periodic boundary conditions on a unit cell $a\times b$.\cite{hierarchy, haldane_torus} This also produces a finite set of $N_\phi=ab/2\pi l_B^2$ single-particle states which are periodic functions under the transformation of coordinates $x\rightarrow x+ a, y \rightarrow y +b$ (we assume that there are no additional phases generated by such a transformation). Because of the presence of magnetic field, the many-body Hamiltonian reduces in the invariant subspace of the magnetic translation group\cite{haldane_torus} and the eigenstates are labeled by the pseudomomentum $\mathbf{k}$, restricted to the Brillouin zone $(2\pi s/a, 2\pi t /b); s,t = 0, ..., \bar N-1,$ where $\bar N$ is the greatest common divisor of $N, N_\phi$. Contrary to the sphere, trial states on the torus are uniquely specified by their filling factor and thus we can directly address transitions between the 331 state, the Pfaffian and the CFL. Moreover, states in this geometry are distinguished by their ground-state degeneracy. If the filling factor is $\nu=p/q$ ($p,q$ having no common divisor), there is a generic degeneracy of $q$ which comes from the center of mass motion\cite{haldane_torus} having no physical importance, so we implicitly assume it in what follows. Apart from this, there can be additional degeneracies occuring due to special point symmetry of the Brillouin zone (trivial) or those that arise from the multicomponent\cite{wz} or the non-Abelian nature of the state.\cite{mr, readrezayi} For the 331 state we expect a quadruplet of states (up to the center of mass degeneracy) one of which belongs to the $\mathbf{k}=0$ sector of the Hilbert space and the remaining three are located at the corners of the Brillouin  zone, $\mathbf{k}=(0,\bar N/2), (\bar N/2, 0), (\bar N/2, \bar N/2)$. In contrast to the 331 state, the Moore-Read Pfaffian has only a threefold degeneracy\cite{read_green} $\mathbf{k}=(0,\bar N/2), (\bar N/2, 0), (\bar N/2, \bar N/2)$, whereas compressible states in general do not possess clearly defined degeneracies (they may appear to have accidental degeneracies which are functions of the aspect ratio of the torus $a/b$, particle number and any other parameter). These are the expectations based on the analytic form of the trial wave functions and their parent conformal field theories,\cite{mr, mvm_nr} but they are also borne out exactly in the numerical diagonalization of the model Hamiltonians. \cite{sheng, rh}

\begin{figure}[htb]
\centering
\includegraphics[width=\textwidth, angle=270, scale=0.32]{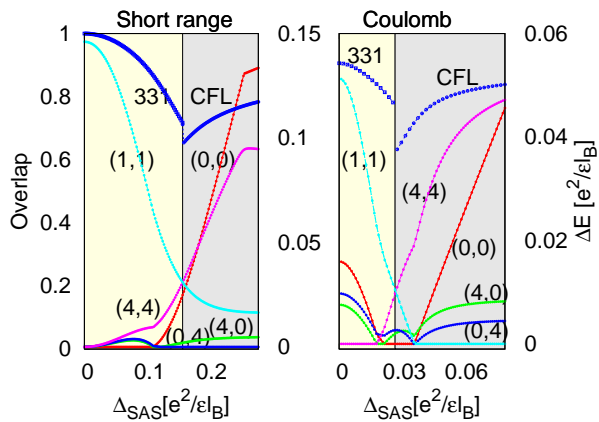}
\caption{(Color online) Low-energy part of the spectrum (relative to the ground state) of the  short-range 331 (left panel) and the long-range Coulomb Hamiltonian (right panel) for $N=8$ electrons on the torus at $\nu=1/2$ and aspect ratio $a/b=0.97$ as a function of tunneling $\Delta_{SAS}$ (right axis). Also shown (left axis) is the overlap with the phases we identify as the 331 state and the CFL.}
\label{fig_torus_331}
\end{figure}

In Fig.\ref{fig_torus_331} we show the relevant low energy part of the spectrum of the Hamiltonians (\ref{shortrange}) and (\ref{longrange}) as a function of tunneling, measured relative to the ground state (right axis), for $N=8$ electrons and the fixed aspect ratio $0.97$ in the vicinity of the square unit cell. We identify the multiplet of 4 states that build up the 331 phase, whose exact degeneracy for the short-range Hamiltonian and small $\Delta_{SAS}$ (left panel) is partially lifted in case of the Coulomb interaction (right panel). 331 phase is destroyed for sufficiently large $\Delta_{SAS}$ when the $\mathbf{k}=(1,1)$ state (fourfold degenerate) comes down and eventually forms a gapless branch with $(0,4)$ and $(4,0)$ members of the 331 multiplet (other excited states not shown). We identify the large-tunneling phase as the CFL phase because exactly the same spectrum is seen in a single layer with Coulomb interaction and the same aspect ratio. This transition is quantitatively reflected also in the overlap with the trial 331 states and CFL as a function of tunneling (Fig. \ref{fig_torus_331}, left axis). In order to take into account the ground state degeneracy, we define overlap on the torus in the following way:
\begin{eqnarray}\label{overlap_torus}
|\langle \Psi_{\mathrm{trial}} | \Psi \rangle |^2 \equiv \frac{1}{|\mathcal{S}_{\mathrm{trial}}|} \sum_{\mathbf{k}\in \mathcal{S}_{\mathrm{trial}}} |\langle \Psi_{\mathrm{trial}}(\mathbf{k}) | \Psi (\mathbf{k}) \rangle|^2 ,
\end{eqnarray}
where $\mathcal{S}_{\mathrm{trial}}$ stands for the degenerate subspace expected for $|\Psi_{\mathrm{trial}}\rangle$. This amounts to adding together the overlap squared for each of the expected members in the ground state multiplet (normalizing the sum by the expected ground state degeneracy $|\mathcal{S}_{\mathrm{trial}}|$ to be $1$ at maximum) and the definition is obviously meaningful only in the case where we had previously established the correct level ordering in the spectrum. 

\begin{figure}[htb]
\centering
\includegraphics[width=\textwidth, angle=270, scale=0.32]{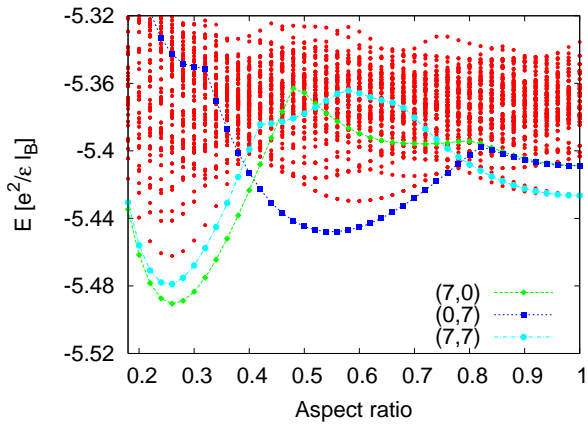}
\caption{(Color online) Spectrum of the $N=14$ electrons in a single layer at $\nu=1/2$ interacting with $V^+(r)$ ($d=l_B$), as a function of aspect ratio. We highlight the states with quantum numbers of the Moore-Read Pfaffian.}
\label{fig_torus_coulomb_spectrum}
\end{figure}

Upon a closer look at Fig. \ref{fig_torus_331}, one notices that the torus spectra suggest little qualitative difference between the short-range and the Coulomb Hamiltonian. In particular, we do not see any indication of the Pfaffian threefold ground-state degeneracy for large $\Delta_{SAS}$ which could be expected from the large overlap on the sphere (Fig. \ref{fig_sphere}). To reconcile these two results, we again focus on the large tunneling limit and vary the aspect ratio of the torus to investigate the possibility of an emergent Pfaffian phase (Fig. \ref{fig_torus_coulomb_spectrum}). We assume that in the large tunneling limit, we have effectively a single layer (polarized) ground state for the symmetric interaction $V^+(r)$. In Fig. \ref{fig_torus_coulomb_spectrum} we show the spectrum of the single layer system of  $N=14$ electrons interacting with $V^+(r)$ as a function of aspect ratio and connect the levels that have the quantum numbers of the Moore-Read Pfaffian. We also include the background charge correction.\cite{yhl} One notices that, with the exception of a very narrow range of aspect ratios around $0.4$, there is no evidence of a clear Moore-Read degeneracy. A narrow region where we see the threefold multiplet of states for $N=14$ also exists for $N=8$, but is obscured by the presence of higher energy levels in systems of $N=10$ and $12$ electrons. Thus we conclude that it cannot represent a stable phase, but a possibility remains that it is a critical phase which becomes stronger as one approaches the thermodynamic limit or as one changes the interaction away from the pure Coulomb.  

We note that varying $d$ (at the fixed aspect ratio) does not lead to any qualitative change in the ground state degeneracy as long as $V^+(r)$ interaction is used. This is clearly different from Zhang-Das Sarma interaction which induces level crossings in the spectrum in such a way that for large $d$ (typically beyond $4l_B$), a Pfaffian degeneracy is seen for big enough systems,\cite{pjds} with the exception of $N=10$. This is similar to the results in the second Landau level\cite{pjds}, as well as the calculations on the sphere\cite{papic_zds}, but the prohibitively small gap suggests that such a state, if it exists, is very fragile. 

\subsection{Pfaffian signatures for intermediate tunneling and a proposal for the phase diagram}
\label{sec_discussion}

\begin{figure}[htb]
\centering
\includegraphics[width=\textwidth, angle=0, scale=0.32]{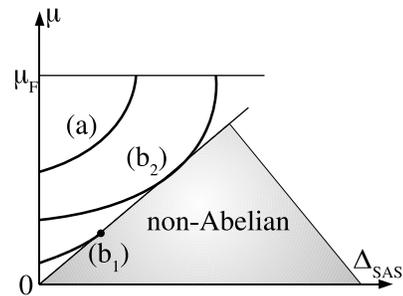}
\caption{Possible outcomes of tunneling on a twocomponent system such as the transition to a Fermi liquid ($a$) or to a critical Moore-Read Pfaffian ($b_1, b_2$), in the context of phase diagram after Read and Green.\cite{read_green} Note that the value for $\mu_F\sim 1/m^*$ is interaction-dependent due to the renormalized CF mass $m^*$ and we may have different dividing lines $\mu=\mu_F$ depending on the kind of the interaction.\cite{hlr}}
\label{fig_diagram}
\end{figure}

We conclude this section with a summary of our exact-diagonalization results in the two geometries in order to make a connection with the BCS analysis of Sec. \ref{sec_tunneling} and sketch possible paths of the $\nu=1/2$ twocomponent system with tunneling in the phase diagram of Read and Green,\cite{read_green} see Fig. \ref{fig_diagram}. In Fig. \ref{fig_diagram}, $\mu$ has the meaning of the effective chemical potential $\mu^{\rm eff}$ of the whole system as in Sec. \ref{sec_tunneling}, renormalized by $\Delta_{SAS}$, i.e. $\mu = \mu (\Delta_{SAS})$. It is assumed that it can be approximated by the value of the chemical potential of the  dominant even channel, $\mu^{\rm eff}\simeq \mu^e$ and the separation between the Abelian and non-Abelian phases in Fig. \ref{fig_diagram} is defined by setting then the value of the chemical potential of the odd channel to zero, i.e. $\mu^o\simeq \mu (\Delta_{SAS}) - \Delta_{SAS} = 0.$ This approximation renders necessary taking into account the renormalization of the parameters in the BCS Hamiltonian (\ref{ham}) with tunneling, as in Ref. \onlinecite{read_green}.

On the sphere, we first recall a very large difference in $\Delta_{SAS}^C$, the critical value of tunneling required to fully polarize the system in the $x$-direction, for the two interactions considered. A much larger value for the short-range 331 Hamiltonian suggests that the chemical potential for the even channel in this case is much more strongly renormalized than for the long-range Coulomb interaction and therefore such a system may assume a phase trajectory labeled $(a)$ in Fig. \ref{fig_diagram}, directly moving from 331 state through the Abelian phase and into a CFL. 

A question we ask at this point is whether the CFL, a likely phase at $P=1$, leaves room for other one-component states to form as we increase the tunneling. In particular, is there a possibility for a system to evolve along the trajectory which terminates at ($b_1$) or touches ($b_2$) the critical line that separates the Abelian from the non-Abelian phase in Fig. \ref{fig_diagram}? Such an intermediate phase could possess significant overlap with the Moore-Read Pfaffian, but it would necessarily have a small gap and we refer to it  as ``critical Pfaffian". 

\begin{figure}[htb]
\centering
\includegraphics[width=\textwidth, angle=270, scale=0.3]{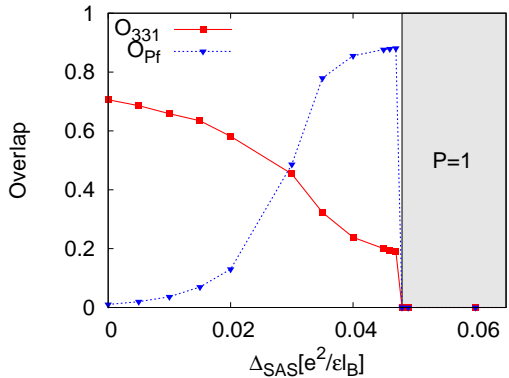}
\caption{(Color online) Overlaps between the exact ground state of $N=10$ electrons on the sphere with the 331 state ($O_{331}$) and the Pfaffian ($O_{\mathrm{Pf}}$), as a function of tunneling $\Delta_{SAS}$, for the Coulomb bilayer Hamiltonian and $d=0.4l_B$.}
\label{fig_criticalphase_sphere}
\end{figure}

On the sphere, a suitable system to detect the signature of the critical Pfaffian is the Coulomb $N=10$ system where the large-tunneling phase is compressible for $d \lesssim 0.5l_B.$\cite{papic_zds} We therefore fix $d=0.4l_B$ and vary $\Delta_{SAS}$ (Fig. \ref{fig_criticalphase_sphere}).
For $\Delta_{SAS}=0$, we are still largely in the 331 phase and for large $\Delta_{SAS}$ we are in the CFL; however, for intermediate tunnelings we see a developing Pfaffian that establishes in a narrow range around $\Delta_{SAS}=0.04 e^2/\epsilon l_B$. Therefore, despite ``weaker" incompressibility for small $\Delta_{SAS}$ and full compressibility for large $\Delta_{SAS}$, for intermediate tunneling we find evidence for the Pfaffian, as suggested by the trajectory $b_2$ in Fig. \ref{fig_diagram}). For larger values of $d$, as we remarked in Sec. \ref{sec_sphere}, we observe saturation of the overlap for Coulomb interaction and the tentative trajectory in that case resembles $b_1$.

\begin{figure}[htb]
\centering
\includegraphics[width=\textwidth, angle=270, scale=0.32]{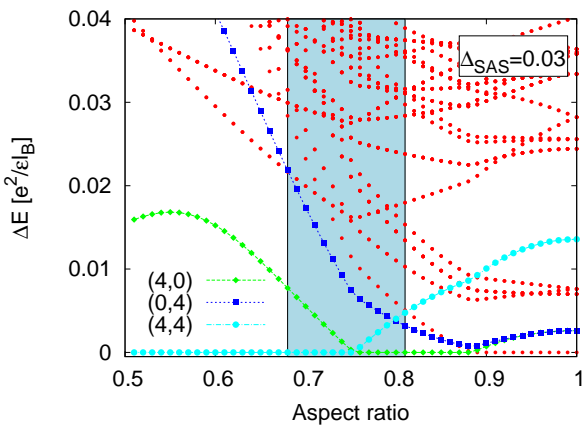}
\caption{(Color online) Low-energy part of the spectrum (relative to the ground state) of the Coulomb Hamiltonian (with $d=l_B$)
for $N=8$ electrons on the torus at $\nu=1/2$ and $\Delta_{SAS}=0.03e^2/\epsilon l_B,$ as a function of the aspect ratio. Shaded region represents the tentative phase with the Pfaffian degeneracy.}
\label{fig_criticalphase_torus}
\end{figure}

The effects of CFL physics are rendered more transparent in the torus geometry, where we have identified the dominant phases as $331$ and CFL (Fig. \ref{fig_torus_331}), with a direct transition between the two of them. 
We choose a value for $\Delta_{SAS}=0.03e^2/\epsilon l_B$ which places the system in the center of the transition region (compare also with Fig. \ref{fig_criticalphase_sphere}) and examine the spectrum of an $N=8$ system as a function of aspect ratio for an emerging Pfaffian degeneracy, Fig. \ref{fig_criticalphase_torus}. 
In torus geometry there is no subtle dependence on $d$, so we take as before $d=l_B$. In agreement with the results on sphere, we find a region of aspect ratios where the correct Pfaffian degeneracy is visible.

Previous results lend support to the scenario of an intermediate critical phase in a long-range Coulomb system, which has a small gap (Fig. \ref{fig_criticalphase_torus}) but possesses large overlap with the Moore-Read Pfaffian (Fig. \ref{fig_criticalphase_sphere}). We would like to stress that all of our conclusions are based on the idealized bilayer Hamiltonian (\ref{main_hamiltonian}). As such, it is not clear at present to what extent they apply to the experiments\cite{luhman, shabani} where the electron density-profile differs significantly from an ideal bilayer. With respect to theoretical studies, stronger indication of topological degeneracy is likely to be found in a model that assumes non-zero thickness of each layer,\cite{pjds} but that would lead also to a substantial decrease of the gap.\cite{storni}

\section{Generalized tunneling constraint}\label{sec_generalizedtunneling}

In Sec. \ref{sec_numerics} we found that in a system with a fixed number of particles and the tunneling commonly expressed as $-\Delta_{SAS}S_x$, there is only circumstantial evidence for the clear Pfaffian phase in finite systems that can be studied by ED. This evidence appears most striking when Coulomb overlaps in the spherical geometry are considered (Fig. \ref{fig_sphere}). However, these overlaps must be interpreted with caution: on the sphere we can indeed only study the competition between Pfaffian and an \emph{excited} CFL state (containing a quasiparticle) because the ground-state candidates occur at different values for the shift $\delta$ as mentioned in Sec. \ref{sec_sphere}. As we have explained there, the latter state is indeed favored by short-range interactions. Moreover, even when the ground state is described by the Pfaffian state, the energy spectrum is not that of a typical incompressible state with a gap to all excitations. Therefore, an explanation for the large Coulomb overlap with the Pfaffian may be the shift which artifically favors it over the CFL phase. This interpretation agrees with the results in the torus geometry, which treats both phases on the same footing and which suggests that the CFL is the likely outcome of tunneling on the 331 state. We can give two general arguments for this. First, strong tunneling has a tendency to polarize the spin in the $x$-direction, and one therefore crosses over to an effective one-component system that in the LLL favors the formation of a compressible CFL phase. Second, we have shown in the BCS approach for the charge-neutral sector (Sec. \ref{sec_tunneling}) that tunneling does not only lower the chemical potential $\mu^o$ in the odd channel but also {\sl increases} that in the even channel, $\mu^e$. We argued that this leads to the insufficiency of the BCS model description, which then describes a local maximum and the system crosses over to a CFL.

It is clear that, in addition to tunneling, one also must find a way to prevent the effective even-channel chemical potential from becoming too large if the weakly-paired phase is to be established in the system.  In this section, we propose a way to implement this requirement formally via generalized tunneling constraint. On the level of the BCS model used in Sec. \ref{sec_tunneling}, this constraint leads to a stable weak-pairing phase in the even channel. In the following subsection A we describe a formal implementation of the constraint in a system of BCS paired fermions. On the basis of this model, we propose, in subsection B a system in contact with a reservoir with which it can exchange particles so that, with the tunneling term included, a stable Pfaffian phase can be reached.

\subsection{A system with generalized tunneling constraint}

Cauchy determinant pairing describes $p$-wave pairing of $\uparrow$ and $\downarrow$ particles, and in order to recover the spinless Pfaffian pairing, we need to ``identify" $\uparrow$ with $\downarrow$. Within a functional formalism, that amounts to adding a term of the form

\begin{equation}
\chi({\bf r})[\Psi_{\uparrow}({\bf r}) - \Psi_{\downarrow}({\bf r})]
\end{equation}
to the Langragian density via Grassmannian Lagrange multiplier $\chi({\bf r})$. We will assume instead that we can alternatively express this via the constraint
\begin{equation}\label{constraint}
\lambda({\bf r})[\Psi_{\uparrow}^{\dagger}({\bf r}) -
\Psi_{\downarrow}^{\dagger}({\bf r})][\Psi_{\uparrow}({\bf r}) -
\Psi_{\downarrow}({\bf r})],
\end{equation}
in terms of the bosonic multiplier $\lambda({\bf r})$. By construction this constraint affects only the odd channel. Within the mean-field approximation of a spatially constant multiplier $\lambda({\bf r})=\lambda$, one may identify $\lambda = \Delta_{SAS}/2$, i.e. the effect of the multiplier is the 
same as the tunneling term in Sec. \ref{sec_tunneling}, except for an overall decrease of the chemical potential, $\mu\rightarrow \mu-\lambda$, which eventually yields a $\lambda$-independent chemical potential in the even channel, $\mu^e=\mu$, as mentioned above. Integration over the Lagrange multiplier projects to $\Psi_{o}^{\dagger} \Psi_{o} = 0$, where $\Psi_{o}^{\dagger} = [\Psi_{\uparrow}^{\dagger}({\bf r}) - 
\Psi_{\downarrow}^{\dagger}({\bf r})]/\sqrt{2}$ is again the odd spin superpostion written in terms of the fermion fields
$\Psi_{\sigma}({\bf r})$, i.e. it leaves us with no density in the odd channel.

The BCS Hamiltonian including the constraint (\ref{constraint}) has the same form as 
(\ref{ham}) except that now $\etilde_{\bk} = \epsilon_{\bk} - \mu + \lambda$, as a consequence of the above-mentioned shift in the chemical potential. We can diagonalize it by a Bogoliubov transformation,
\begin{equation}
\alpha_{\bk} = u_{\uparrow} c_{\bk \uparrow} + u_{\downarrow} c_{\bk
\downarrow} + v_{\uparrow} c_{-\bk \uparrow}^{\dagger} +
v_{\downarrow} c_{-\bk \downarrow}^{\dagger}.
\end{equation}
The equation $[\alpha_{\bk},H] = E \alpha_{\bk}$ then defines the Bogoliubov-de Gennes equations and the Hamiltonian is transformed into the canonical form $H = \sum_{\bk} E_{\alpha \bk} \alpha_{\bk}^{\dagger} \alpha_{\bk}+ \sum_{\bk}
E_{\beta \bk} \beta_{\bk}^{\dagger} \beta_{\bk}+\tilde{E}_{0},$ where the eigenvalues $\pm E_{\alpha\bk}$ and $\pm E_{\beta\bk}$ 
are given by
\begin{eqnarray}\label{eigenvalues} 
&& E_{\alpha\bk} = \sqrt{\epsilon_{\bk,1}^{2} + \Delta_0^{2}k^2}\;\; {\rm ,\,\, with}\;\;
\epsilon_{\bk,1} =
\epsilon_{\bk} - \mu \nonumber, \\
\nonumber
&& E_{\beta \bk} = \sqrt{\epsilon_{\bk,2}^{2} + \Delta_0^{2}k^2}\;\; {\rm ,\,\, with} \;\;
\epsilon_{\bk,2} = \epsilon_{\bk} - \mu + 2 \lambda,\\
\end{eqnarray}
respectively.
The eigenvectors corresponding to $E_{\alpha\bk}$ and $E_{\beta\bk}$ are, respectively,
\begin{eqnarray}\label{alphas}
\nonumber \alpha_{\bk} &=& \frac{k}{2 \sqrt{E_{\alpha\bk}}} \frac{\sqrt{E_{\alpha\bk} + \epsilon_{\bk,1}}}{k_x-ik_y}  (
c_{\bk \uparrow} + c_{\bk \downarrow} ) \\
&& +\frac{\Delta_0 k}{2 \sqrt{E_{\alpha \bk}}} \frac{1}{\sqrt{E_{\alpha\bk} + \epsilon_{\bk,1}}} ( c_{-\bk \uparrow}^{\dagger}+
 c_{-\bk \downarrow}^{\dagger})
\nonumber \\
\nonumber \beta_{\bk} &=& \frac{k}{2 \sqrt{E_{\beta\bk}}}\frac{\sqrt{E_{\beta\bk} + \epsilon_{\bk,2}}}{k_x-ik_y} (
 c_{\bk \uparrow} - c_{\bk \downarrow} ) \\
&& -\frac{\Delta_0k}{2 \sqrt{E_{\beta\bk}}} \frac{1}{\sqrt{E_{\beta} + \epsilon_{\bk,2}}} (c_{-\bk \uparrow}^{\dagger} - c_{-\bk\downarrow}^{\dagger}).
\end{eqnarray}
To find the stationary point for the action defined by the diagonalized BCS Hamiltonian at zero temperature, it is useful to continue $\lambda$ from the real axis to the complex plane $\field{C}$ (see Ref. \onlinecite{assa} for details). The part $\tilde{E}_0$, through its dependence on $\lambda$,

\begin{equation}
\tilde{E}_{0} = - \sum_{\bk} \frac{\Delta_0^{2}k^2}{2} \left( \frac{1}{E_{\alpha\bk} +
\epsilon_{\bk,1}} + \frac{1}{E_{\beta\bk} +
\epsilon_{\bk,2}}\right),
\end{equation}
determines the stationary point: $\partial \tilde{E}_{0}/\partial \lambda = 0.$ Introducing the notation $\mu - 2 \lambda \equiv \tilde{\mu} \in \field{C}$ for the analytically continued chemical potential in the odd channel,
we continue as $E_{\beta\bk} + \epsilon_{\bk,2} = |\tilde{\mu}| - \tilde{\mu} + o(k^{2}),$ which gives us the stationary point ${\rm Re}\tilde{\mu} \rightarrow - \infty $ or, equivalently, $\lambda \rightarrow + \infty$ and we have strong coupling in the odd channel. 
For our choice $\mu >0$, we thus obtain weak coupling in the even channel that has a Pfaffian description\cite{read_green} 
in $c_{\bk \uparrow} + c_{\bk\downarrow}$ (even) variables [see the expression for $\alpha_{\bk}$ (\ref{alphas})].
Unlike the situation in Sec. \ref{sec_tunneling}, with the same assumption that $\Delta_0$ does not renormalize strongly with tunneling, the chemical potential of the even channel stays the same and the even channel can thus be weakly paired.

\subsection{A system with fluctuating density}

The previous discussion was on the simplified model of the 331 physics in
terms of neutral fermions with an additional constraint that leads
to the Pfaffian, but seems artificial and hard to implement in an
experimental setting. Nevertheless it suggests a possible way to
achieve the stable Pfaffian phase. With respect to ordinary
tunneling the generalized constraint can be modeled by strong tunneling
and an additional term in the effective description, $ \lambda N$, where
$N$ is the total number of particles and $\lambda$ tunneling strength
as before.\cite{fnote}
The chemical potential (of the whole system)
depends on the tunneling and changes as $\mu - \lambda$. Then we have
the following physical picture in mind: as we take $ \lambda > 0$
this decrease of the chemical potential with tunneling will
imply the decrease of the density of the system. On the other
hand, from the solution of the BCS system with the
generalized constraint, we see that the effective chemical
potential of the even channel stays the same - equal to $\mu$, Eq. (\ref{eigenvalues}). 
This means also that the number of particles
in the even channel stays the same, so the effects of the
tunneling and the additional term cancels, but the polarization $P$
increases with tunneling. Thus we effectively maintain the
same effective parameter $\mu$ with tunneling, its value will
not increase, and we will be able to achieve the stable Pfaffian
phase. To be more specific and
quantitative about the role of fluctuating density in achieving
the Pfaffian, we discuss in the remainder of this section the necessary dependence of density
on tunneling. We will recover the demand
for the decreasing density for the case of strong tunneling.

\begin{figure}[htb]
\centering
\includegraphics[width=\textwidth, angle=0, scale=0.32]{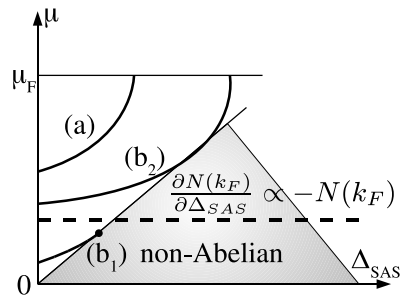}
\caption{Generalized tunneling contraint (dashed line) in the phase diagram of Fig. \ref{fig_diagram}. Note that, as in Fig. \ref{fig_diagram}, $\mu$ denotes the effective chemical potential equal to the chemical potential of the even channel in the strong tunneling limit.}
\label{fig_diagram2}
\end{figure}

Our BCS Hamiltonian, which is also the thermodynamic potential $\Omega$ at zero temperature, is specific for the fact that the independent thermodynamic variable, along with volume and temperature, is the chemical potential $\mu^\lambda$, given by $ \mu^\lambda = \mu - \lambda$. Therefore, as we change $\lambda$ (the parameter of the Hamiltonian), we also induce a change in the chemical potential $\delta \mu^\lambda = - \lambda$ and this implies
\begin{equation}\label{condition_eq}
\frac{\partial \Omega}{\partial \lambda} = \frac{\partial \tilde{E}_{0}}{\partial \lambda} = N,
\end{equation}
for the particular system. The effective chemical potential, on the other hand, is in this case
\begin{equation}\label{mutot}
\mu^{\rm eff} = \mu P + (\mu - 2 \lambda) (1 - P) = \mu + (P - 1)\; 2
\lambda.
\end{equation}
As before, this relation shows how the chemical potential renormalizes as a parameter $\lambda$ of the BCS Hamiltonian is varied.
We keep the volume constant and measure $N$ with $k_{F}$ in the usual ansatz:
\begin{equation}
N(k_{F}) = \sum_{\bk,\sigma} \rightarrow \frac{2 \times 2\pi}{(2\pi)^2} \int_{0}^{k_{F}}dk k = k_{F}^{2}/2\pi,
\end{equation}
and Eq. (\ref{condition_eq}) becomes
\begin{equation}
N(k_{F}) = \frac{\partial \tilde{E}_{0}(k_{F},\lambda)}{\partial
k_{F}} \frac{\partial k_{F}}{\partial \lambda} + \frac{\partial
\tilde{E}_{0}(k_{F},\lambda)}{\partial \lambda}. \label{eq}
\end{equation}
Differentiating and converting the sum in $\tilde{E}_0$ into an integral over $k$, we obtain the partial differential equation,
\begin{eqnarray}\label{bla}
\nonumber k_{F}^{2} &=& - \frac{\Delta_{0}^{2} k_{F}^{3}}{2}
\left(\frac{1}{E_{\alpha \bk_F} + \epsilon_{\bk_F,1}} + \frac{1}{E_{\beta\bk_F} +
\epsilon_{\bk_F,2}}\right) \frac{\partial k_{F}}{\partial \lambda} \\
&& +\int_{0}^{k_{F}}dk\; k^{3} \frac{\Delta_{0}^{2}}{E_{\beta \bk} (E_{\beta \bk}
+ \epsilon_{\bk,2})},
\end{eqnarray}
from which we can extract the limiting case $\lambda \gg \hbar^2 k_{F}^{2}/2m^*$ when
\begin{equation}\label{bla2}
\frac{\partial N(k_{F})}{\partial \lambda} = - c N(k_{F}),
\end{equation}
with the constant factor $c > 0 $, i.e. for large tunneling we should decrease the
density of the system to stay at small $\mu^{e}$ and to stabilize the Pfaffian.  
We expect this limit to be pertinent to the experiments such as Ref. \onlinecite{shabani} where the compressible state starts to show signs of incompressibility upon density imbalance. Enforcing the
condition (\ref{bla2}) is then expected to lead to strengthening of the paired Pfaffian state, as summarized in Fig. \ref{fig_diagram2} where
the dashed line represents schematically a phase trajectory of the system that evolves under the generalized tunneling constraint (in the large $\Delta_{SAS}$ limit). As Eq. (\ref{bla2}) shows,
in this case the density of the system needs to be decreased simultaneously with the increase of tunneling $\Delta_{SAS}$. This can be achieved
conventionally via the application of a gate voltage to the evaporated top/bottom gates as e.g. in Refs. \onlinecite{csathy, shabani} or by growing \emph{in situ} a $n^+$-GaAs layer that can serve as a gate. \cite{nuebler}  

The main reason why we find it desirable to have an open system in the experiment
is our inability to specify the parameters of the simplified, effective
Read-Green model and pin-point the optimal density for which the Pfaffian
state would be strong. With a better knowledge of the microscopic details of the system,
one may as well fix a particular density (which we expect to be small) for a given, not too large,
tunneling strength where the Pfaffian will be stable. However, an open system would allow a systematic
study over a range of tunneling strengths in which the Pfaffian would show a characteristic 
strengthening that would distinguish it from any other incompressible candidates.

Therefore, in principle, by changing the density of the system we can achieve a stable Pfaffian phase. 
We would like to compare the present discussion which is based on the simplified model of neutral fermions (mean field
in nature with the simplifying assumption $\Delta_{o} = - \Delta_{e} = {\rm constant}$ i.e. independent of tunneling) with
exact diagonalization in the LLL in the previous section. The arguments presented here call for an open system
with adjustable density which demands also the adjustment of the magnetic field $B$ to achieve the stable Pfaffian phase. 
Although doing so will preserve the filling factor,  in general changing the total density may enhance the role of LL mixing and thus invalidate the LLL assumption in the exact diagonalizations. It will also lead to the renormalization of the parameters of the BCS effective model not taken into account in Sec. \ref{sec_tunneling}. Indeed, when lowering the total chemical potential via the generalized constraint, the density is also decreased. On the other hand the ratio between interaction strength $(e^{2}/l_{B})$ and cyclotron frequency $ (\omega_{c} \sim 1/l_{B}^{2})$ is proportional to $ 1/\sqrt{\rho}$, from which we see that the LLL projection is invalidated if the density is significantly reduced. Therefore to reach and establish the Pfaffian phase it is likely that LL mixing has to be taken into account. This has been discussed in the recent literature \cite{SimonRezayi,BisharaNayak} as a way to stabilize the Pfaffian phase. Here we seek the Pfaffian in a two-component setting when
a parameter of the system $\lambda$ is varied, which makes the inclusion of higher LLs harder. 
If we remain in the LLL, changing of the density amounts to simple rescaling of spectra $(e^{2}/l_{B}\rightarrow
c\; e^{2}/l_{B}$ with $c > 0$), which cannot induce any signficant effect such as the change in the nature of (quasi)degeneracy
of ground states on the torus.  Even if the evidence for a Pfaffian phase is rather weak for the system sizes studied here, 
one may hope that the increase of these sizes will improve the case for such a state in the
LLL as the odd channel may assume the role of the first excited Landau level before a complete polarization.

\section{Conclusions}\label{sec_conclusions}

We investigated the possibility of creating the Moore-Read Pfaffian out of the paired two-component 331 state via tunneling. Exact diagonalization, performed under the constraint of LLL projection and fixed total number of particles, could not detect a stable Pfaffian phase, but a critical one between 331 and CFL phases. While the short-range interaction is likely to favor a direct transition from the 331 to the CFL phase, long-range Coulomb interactions leave the possibility for a Pfaffian-like phase if the parameters of the system are tuned in a special way. Based on the connection between our numerical results and the effective BCS Hamiltonian theory of paired states, we argue that one way to stabilize the Pfaffian state is to change the density (number of particles) of the system while increasing the tunneling.

Our analysis was restricted to the Coulomb bilayer system and the tunneling term of the form $-\Delta_{SAS}S_x$, which is small in magnitude and generally difficult to control. Similar considerations apply to the quantum well systems\cite{shabani} where $\uparrow,\downarrow$ stand for the two lowest electronic subbands and $\Delta_{SAS}$ acts like a Zeeman energy. In these systems, the analogue of the tunneling term used in our paper can be deployed to create asymmetric charge distribution in the wide quantum well.\cite{shabani} The interplay of these two kinds of terms, tunneling and Zeeman, will be addressed in future work. 

\section*{Acknowledgments}
This work was supported by the Serbian Ministry of Science under Grant No. 141035, by the Agence Nationale de la Recherche under Grant No. ANR-JCJC-0003-01 and by grants from R\'egion Ile-de-France. Illuminating discussions with H. Hansson, S. Das Sarma and M. Shayegan are gratefully acknowledged. ZP acknowledges discussions with D. Sheng, D. Yoshioka and especially Th. Jolicoeur and K. Vyborny.


\begin{thebibliography}{9}
\bibitem{fnote} This is equivalent to imposing the following constraint for the BCS Hamiltonian (\ref{ham}) with only a tunneling term $\Delta_{SAS}$: the increase of the tunneling $\Delta_{SAS}$ is followed by the decrease in the chemical potential $\mu$ by twice the
same amount, $\delta \mu + \delta \Delta_{SAS}/2 = 0$, in our simplified model to maintain the chemical potential
of the even channel, $\mu^e=\mu + \Delta_{SAS}/2$, stay constant.
\end{thebibliography}
\end{document}